\documentclass[a4paper,12pt]{article}
\usepackage{times,amsmath,amssymb}
\usepackage{here,cite,fancyhdr,enumerate}
\usepackage[margin=25mm]{geometry}
\makeatletter
\renewcommand{\section}{\@startsection{section}{1}{\z@}%
                                     {-3.25ex\@plus -1ex \@minus -.2ex}%
                                     {1.5ex \@plus .2ex}%
                                     {\textnormal\normalsize\bfseries}}
\renewcommand{\subsection}{\@startsection{subsection}{2}{\z@}%
                                     {-3.25ex\@plus -1ex \@minus -.2ex}%
                                     {1.5ex \@plus .2ex}%
                                     {\textnormal\normalsize\bfseries}}
\makeatother
\begin{document}
% CHANGE Topic Acronym ------------------- vvvvv
\fancyhead[L]{\textit{Topic area:}~\textbf{NUPP}}
\fancyfoot[L]{\textit{Presenting author's name:}\hspace*{2em}%
% CHANGE presenting author's name in the line below
\textbf{R.~Delbourgo}}
\begin{raggedright}
% CHANGE paper title in the line below
\section*{\large Why Three Generations?}
% CHANGE author list and affiliations in the lines below
\underline{R. Delbourgo}$^{1}$ \\
\vspace*{1ex}\ \\
$^{1}$\textit{School of Mathematics and Physics, University of Tasmania,
GPO Box 252-21, Hobart 7001}\\
\vspace*{0.5ex}\ \\

% CHANGE email address in the line below - vvvvvvvvvvvvvvvv
\textit{e-mail of corresponding author:}  Bob.Delbourgo@utas.edu.au
%
% Replace the following text (down to 'end Author input') with your own 
% manuscript. Pay particular attention to the method of citation.
%
\subsection*{Introduction}
Anticommuting operators and fields have been known for a long time in quantum
theory but only since the early 70s have anticommuting {\em a-numbers} played a 
prominent role in physics, when supersymmetry (SUSY) and BRST symmetry came to
the fore. (Of course Grassmann numbers were invented much earlier in mathematics
and feature prominently in differential geometry.) Since the 70s they have 
dominated attempts to construct a sensible, properly unified theory.

While BRST symmetry has simplified the proof of the renormalizability of QCD 
and assists the development of gauge theory, the applicability of SUSY has 
turned out to be much more problematic. Nature seems to exhibit no sign of 
supersymmetric partners at our present energy scales and yet this has not
prevented a large numbers of researchers devoting all their time and effort into
developing the subject, both in the context of ordinary field theory and
string/brane extensions. Apart from the alluring beauty of SUSY, there are 
basically two reasons for this: it is the only consistent higher symmetry theory 
that combines the Poincar\'e Group with internal symmetries in a nontrivial way, 
and it naturally leads to the cancellation of quantum infinities between bosons 
and fermion contributions without any fine-tuning. Without jeopardising these
theoretical successes of SUSY, I would like to outline a scheme which uses 
a-numbers in a different way agreeing with the known particle spectrum and 
the character of their interactions, and which may lead to the goal of a unified 
theory.

The principle behind this scheme is cancellation of dimensions between c-nos 
and a-nos. It is rather well-known that Bose and Fermi states contribute 
oppositely to statistical formulae, functional determinants and group 
properties\cite{mckhr}; 
indeed this is the fundamental reason for the cancellation between bosonic and
fermionic quantum loops. Because the numbers of visible, commuting space-time 
coordinates $x$ is just 4, I propose to append 5 complex anticommuting 
coordinates $\xi$ to these (but will only use half of the superfield expansions
in $\xi$, making it equivalent to four a-numbers) in order to ensure total
zero-dimensionality of the universe, as it was before the big bang one presumes?
I will associate these a-nos. with `properties' or internal structure, 
giving me a theory of `spacetime-property'; not only will we know where and when
an event occurs, but what it is! A welcome bonus of the scheme is how 
it naturally accommodates three generations, without invoking a particular 
gauge group---normally conjured out of thin air---or repetition number. Further
it mimics Klein-Kaluza type models without producing $\infty$ numbers of modes, 
because a-number expansions are necessarily finite. All of these considerations
suggest that a-number extensions to spacetime may be the way to go, rather than 
higher bosonic coordinates with their infinite excitations, compactification 
notwithstanding. Due to space limitations I shall sketch out the main
ideas and refer you to earlier papers which I have written with various
valued collaborators\cite{djw,dtz}. To my knowledge the only venture into the 
a-no. `property market' besides ours is that of Ellicott and Toms\cite{et}.

%Proceedings papers have a maximum limit of four (4) pages. 

\subsection*{Superfield expansions}
It turns out that three or fewer $\xi$ are not enough to accommodate three
generations. Four $\xi$ suffice, but at the price of `schizosymmetry'
\cite{fj}, causing some discomfort because standard statistics must be 
interpreted unconventionally.
The easiest solution is to add an extra (uncharged) a-number $\xi_0$ to the 
$\xi_\mu$, $\mu = 1...4$, but only consider superfields with odd powers of 
all $\xi$ for fermions $\Psi_\alpha$, even powers for bosons $\Phi$. 
This patches the statistics and leaves us with {\em effectively four} $\xi$.
Taking a leaf out of SU(5) unified gauge theory, the charge and fermion
number assignments are respectively taken to be 
$Q(\xi_0,\xi_1,\xi_2,\xi_3,\xi_4)$ = 0,1/3,1/3,1/3,-1  and 
$F(\xi_0,\xi_1,\xi_2,\xi_3,\xi_4)$ = 1,-1/3,-1/3,-1/3,1. 
Note in passing that Tr($Q$) = 0 helps with anomaly cancellation. 
[The $\xi$ are complex and given by the combination $\xi_{(1)}+i\xi_{(2)}$ of 
the more familiar symplectic basis $\xi_{(1,2)}$.] Any superfield is to be 
expanded in terms of $\xi_\mu$ {\em and} $\bar{\xi}^\nu$. The $\xi$ and
$\bar{\xi}$ being anticommuting, such expansions end at the fifth power:
$$\Phi(X)\equiv \Phi(x,\xi,\bar{\xi})=\!\!\!\sum_{even~r+\bar{r}}\!\!\!
  (\bar{\xi})^{\bar{r}}\phi_{(\bar{r}),(r)}(x)(\xi)^r; \,
  \Psi_\alpha(X)\equiv \Psi(x,\xi,\bar{\xi})=\!\!\!\sum_{odd~r+\bar{r}}\!\!\!
  (\bar{\xi})^{\bar{r}}\psi_{\alpha(\bar{r}),(r)}(x)(\xi)^r. $$
Roughly speaking we may associate labels $i=1 - 3$ with (down) chromicity
and 4 with charged leptonicity, while 0 corresponds to neutrinocity. However 
bear in mind that products of $\xi$ and $\bar{\xi}$ can lead to other properties
like generation number and (up)colour. This is most readily seen by drawing up 
the particle content in a magic 6$\times$6 square corresponding to $r,\bar{r}$ 
running from 0 to 5. $\Psi,\Phi$ are super-self-conjugate in as much as 
$\psi_{(r),(\bar{r})} = \psi^{(c)}_{(\bar{r}),(r)}$, thereby halving the
number of components, but this still leaves too many (256) components
for comfort. It pays to further halve the field by invoking anti-duality,
$\psi_{(r),(\bar{r})} \sim -\psi_{(5-\bar{r}),(5-r)}$, without affecting 
charge assignments and cutting the number down to 120 states. Considering
the source field $\Psi$, the resulting square contains the following varieties 
of up ($U$), down ($D$), charged lepton ($L$) and neutrinos ($N$), where the 
subscript distinguishes between repetitions. In the table below,
 - are duals, * are conjugates:
\begin{center}
 \begin{tabular}{|l||c|c|c|c|c|c|}  \hline
   $r\backslash\bar{r}$ & 0 & 1 & 2 & 3 & 4 & 5 \\
  \hline \hline
   0 &   & $L_1,N_1,D_1^c$ &         & $L_5^c,D_5,U_1$ &      &   \\
   1 & * &           & $L_{2,3},N_{2,3},D^c_{2,3},U_2$ &  & $L_6,D_6,U_3$&   \\
   2 &   &     *     &          & $L_4,N_4,D^c_{4,7}$&      & - \\
   3 & * &           &    *     &        &  -   &   \\
   4 &   &     *     &          &   *    &      & - \\
   5 & * &           &    *     &        &   *  &    \\
 \hline 
 \end{tabular}
\end{center}

Antiduality eliminates the neutrino state corresponding to the product of all
five $\xi$ as well as doubly charged leptons associated with 
$\bar{\xi}^4\xi_0\xi_1\xi_2\xi_3$ and $\bar{\xi}^4\bar{\xi}_0\xi_1\xi_2\xi_3$.
Notice that the table has three $U$ generations, but more generations of $D$, 
$N$ and $L$ are indicated. Who can be truly sure that they will not be found 
with larger masses? (A sterile neutrino may play a useful role anyway.)
{\em The main point is that their number is finite and small.}
Of greater interest is the occurrence of exotic quarks with charges 4/3 and 5/3,
tied with properties $\bar{\xi}^4\xi_0\xi_i$, $\bar{\xi}^4\xi_i\xi_j$; instead
of being a cause for despair one can speculate that they or the extra $D$ may be
connected with new composite hadrons like $\Theta^+$, as were discovered during 
the last two years. Until the mass matrices of these states are constructed in 
a realistic way all avenues are open.

The Higgs field ought to correspond to the superfield $\Phi$ with its even $\xi$
expansion. Vacuum expectation values must be colour singlets and uncharged; a 
plethora of them is available: we can allow for one $\phi_{(0)(0)}$, one
$\phi_{(0),(4)}$, three $\phi_{(1),(1)}$, four $\phi_{(2),(2)}$ --- with the
understanding that their duals are not independent. These must be able to account
for {\em all} the quark, lepton and neutrino masses however, which is a very
strong constraint! 

To show how this might pan out, let me consider a simplified model in 2D 
space-time with two complex $\xi$, having the properties of `electronicity' and
`protonicity' so their charges and fermion numbers are $Q(\xi_1,\xi_2)$ = -1, 1 
and $F(\xi_1,\xi_2)$ = 1, 1 respectively. In this case invoke superfield 
self-duality to exorcise the doubly charged state $\bar{\xi}^1\xi_2$ as well as 
the `atomic' product property $\xi_1\xi_2$. Assuming self-duality the expansions
read
$$\Psi(\bar{\xi},\xi)=(\bar{B}^m\xi_m+\bar{\xi}^mB_m)(1+\bar{\xi}^n\xi_n)/2,
 \qquad \Phi(\bar{\xi},\xi)=(A+S\bar{\xi}^m\xi_m)(1+\bar{\xi}^n\xi_n)/2 .$$
In this model, upon integrating out the properties via $\int d^2\xi d^2\bar{\xi}$, 
for the typical Lagrangian  ${\cal L} = h\Phi^2\langle \Phi \rangle +
    g\bar{\Psi}\Psi\langle\Phi\rangle$, 
the nonvanishing expectation values $a$ and $s$ produce mass terms:
$$[h(2A^2+S^2) + 2g\bar{B}^nB_n]a + [3hAS + g\bar{B}^nB_n]s.$$
For the above we deduce the three masses $M_{B\pm} = [3M\pm\sqrt{M^2+4m^2}]/2$ 
for bosons and $M_F = m+M$ for fermions.

In 5D the situation is much more complicated because there are many more
expectation values to be accounted for; it may turn out to be quite difficult
to fit all known particle masses with the nine allowable expectation values
$\langle \phi \rangle$. What is certain is that the scheme is quite different 
from the standard simple picture with its two sets of $3\times 3$ mass
matrices (neutrinos and quarks), which has created a feeling of complacency
in the particle physics community.
Hints from physics in respect of new multiquark composites and our inability
to provide a convincing picture of masses and $V_{ij}$ mixings, both in the
quark and neutrino sectors, are indications that not everything is understood
or even 100\% correct in the standard description.

\subsection*{Generalized Relativity}
Where do the gauge fields fit into this picture? One way to introduce them
would be to mimic
SUSY and supergauge the massless free Lagrangian for $\Psi$, but without
added complications of spin. [This would need to be done so that anomalies 
cancel and the number of fermion fields match those of the bosons.] But there 
is a more compelling way, which has the benefit of incorporating gravity. The 
idea is to devise a fermionic version of the familiar Kaluza-Klein (KK) scheme,
without the need for infinite modes that arise from shrinking the usual
additional bosonic coordinates. The method has been published elsewhere\cite{dtz}
and, with minor improvements, I would like to highlight the main points. 
Let $X=(x,\xi)$ stand for the spacetime-property manifold.
\begin{itemize}
\item If we are to use a generalized metric for $X$, one must introduce a 
  fundamental length scale to match $x$ and $\xi$, since properties are
  dimensionless. This presents an opportunity to introduce a fundamental 
  length into physics; of course the gravity scale $\kappa = \sqrt{8\pi G_N}$ 
  is a natural choice, particularly as it enters the spacetime sector.
\item Gravity (plus gauge field products, as we shall see) fall within the 
  $x-x$ sector, gauge fields are accommodated in the $x-\xi$ sector, and one 
  presumes that the Higgs scalars $\Phi$ form a matrix in the $\xi-\xi$ sector.
\item Gauge invariance is connected with the numbers of $\xi$, so the
  full gauge group would have to be SU(5) {\em or perhaps a subgroup}.
\item There is no natural place for a gravitino as the $\xi$ are Lorentz scalar.
\end{itemize}

We envisage a real metric: $ds^2 = dx^m G_{mn} dx^n + dx^m {G_m}^\nu d\xi_\nu
 +d\bar{\xi}^\mu G_{\mu n}dx^n + d\bar{\xi}^\mu {G_\mu}^\nu d\xi_\nu$ where
the tangent space limit corresponds to Minkowskian $G_{ab}
\rightarrow I_{ab}= \eta_{ab}$, ${G_\alpha}^\beta \rightarrow
 {I_\alpha}^\beta = \kappa^2{\delta_\alpha}^\beta$, multiplied at least by 
$(\bar{\xi}\xi)^5$ to ensure that the property integration causes no harm. 
Proceeding to curved space the components should contain the force fields, 
leading one to a `superbein'
$${\bar{E}_M}^A = \left( \begin{array}{cc}
                   {e_m}^a & i\kappa\bar{\xi}^\gamma {A_m}_\gamma^\alpha \\
                    0      & \kappa^2\Phi_\mu^\alpha \end{array} \right), $$
and the metric "tensor"
$$ G_{MN} = {\bar{E}_M}^A I_{AB} {E_N}^B =
            \left( \begin{array}{cc} e^a_me_{an}
 +\kappa^2\bar{\xi}^\gamma {A_m}_\gamma^\alpha{A_n}_\alpha^\delta\xi_\delta
 & \,\,\,i\kappa\bar{\xi}^\gamma {A_m}_\gamma^\alpha\Phi_\alpha^\nu \\ 
 -i\kappa\Phi_\mu^\alpha{A_n}_\alpha^\delta\xi_\delta
    &\kappa^2\Phi_\mu^\alpha\Phi_\alpha^\nu \end{array} \right).$$
One can then show that  the generalized connection contains the curl of the 
gauge field $A$, namely $F = \partial A + A\wedge A$, plus the purely 
gravitational connection, resulting in the generalised anti-self-dual scalar 
curvature, $$R=(R^{(g)}+ \kappa^2{F_{mn}}_\mu^\nu{F^{mn}}_\nu^\mu)
   (1-(\bar{\xi}\xi)^5)/4,$$
as desired, if we disregard the matrix structure of $\Phi$ and 
simply assume a flat expectation value $\Phi = 1$. 

Gauge symmetry corresponds to the special change $\xi_\mu\rightarrow
\xi'_\mu=[\exp(i\Lambda(x))]_\mu^\nu \xi_\nu,\, x \rightarrow x'=x$. Given the 
standard transformation law
$$G_{m\mu}(X) = \frac{\partial X'^K}{\partial x^m}
                \frac{\partial X'^L}{\partial\bar{\xi}^\mu}G'_{KL}
              = \frac{\partial \xi'_\kappa}{\partial x^m}
    \frac{\partial \bar{\xi}'^\lambda}{\partial\bar{\xi}^\mu}G'^\kappa_\lambda
 + \frac{\partial \bar{\xi}'^\lambda}{\partial\bar{\xi}^\mu}G'_{m\lambda},$$
this translates into the usual gauge variation (a matrix in property space),
$$A_m(x) =\exp(-i\Lambda(x))[A'_m(x) - i\partial_m]\exp(i\Lambda(x)).$$
Perhaps an easier way to see this is by writing this particular metric 
length$^2$ in the form,
$$ds^2 = dx^mg_{mn}^{(g)}dx^n+\kappa^2(idx^m\bar{\xi}^\mu{A_m}_\mu^\rho
   +d\bar{\xi}^\rho)(-i{A_n}_\rho^\nu\xi_\nu dx^n + d\xi_\rho).   $$
This argument does not fix what (sub)group is to be gauged in property space
although one most certainly expects to handle the nonabelian colour group and 
the abelian electromagnetic group, to agree with physics. In fact the 
correct choice may well be tied to expectation values of $\Phi$ in the 
property sector, which I happened to set equal to unity above for simplicity.

My exposition has been necessarily sketchy due to lack of space but I trust
that the general ideas have come across. A more detailed version of this 
scheme will be published elsewhere.
\vspace{-0.5cm}

%Your text references \cite{ein04}, \cite{end03}.

%\subsection*{Author check list}
%Make sure that you have inserted:\vspace*{-2ex}\ \\
%\begin{enumerate}[(1)]
%\setlength{\itemsep}{-1ex}
%\item Topic Area acronym
%\item Paper title
%\item Author names and affiliations
%\end{enumerate}

%Save in pdf format with the filename as outlined below:

%Last name of corresponding author\_Initials\_AIP\_Topic Acronym CDX$^*$
%\begin{center}example: Smith\_JF\_AIP\_CMMSP\_CD1\end{center}
%$^*$denotes different abstracts with the same corresponding author in the same
%Topic Area by X $=$ 1 or 2 or 3 etc.

%All manuscripts have to be submitted as PDF files. 

\vspace*{-10ex}\ \\ % bad hack to remove blank space above reference list
\renewcommand{\refname}{}

% end of Author input
\end{raggedright}
\end{document}